\documentclass[showpacs,amsmath,amssymb,aps,twocolumn,pre,superscriptaddress,balancelastpage]{revtex4-1}
\usepackage{graphicx}
\usepackage{color}
\usepackage{bm}
\usepackage{epsfig}
\usepackage{subfigure}
\usepackage{amsmath}
\usepackage{amssymb}
\usepackage[normalem]{ulem}
\usepackage{cancel}
\usepackage{hyperref}
\begin{document}

\title[Role of HOIs MI of BEC with OL potential]{Role of higher-order interactions on the modulational instability of Bose-Einstein condensate trapped in a periodic optical lattice}

\author{S Sabari}
\email{ssabari01@gmail.com}
\affiliation{Institute of Theoretical Physics, UNESP -- Universidade Estadual Paulista, 01156-970 S\~{a}o Paulo, SP, Brazil.}
\author{OT Lekeufack}
\email{lekeufackolivier@gmail.com}
\affiliation{Pure physics Laboratory: Group of nonlinear physics and complex systems, Department of Physics, Faculty of Science, University of Douala, P.O. Box 24157, Douala, Cameroon}
\affiliation{Centre d'Excellence Africain des Technologies de l'Information et de la Communication (CETIC), Universit\'e de Yaound\'e I, Yaound\'e, Cameroun}
\author{SB Yamgoue}
\email{sergebruno@yahoo.fr}
\affiliation{Department of Physics, University of Bamenda, P.O Box 39 Bamenda, Cameroon}
\author{R Tamilthiruvalluvar}
\email{taphy1906@gmail.com}
\affiliation{Centre for Nonlinear Science(CeNSc), Department of Physics, Government College for Women(Autonomous), Kumbakonam – 612001, India}
\author{R Radha}
\email{vittal.cnls@gmail.com}
\affiliation{Centre for Nonlinear Science(CeNSc), Department of Physics, Government College for Women(Autonomous), Kumbakonam – 612001, India}
\date{\today}
\begin{abstract}
In this paper, we investigate the impact of  higher-order interactions on the modulational instability (MI) of Bose-Einstein Condensates (BECs) immersed in an optical lattice potential. We derive the new variational equations for the time evolution of amplitude, phase of modulational perturbation, and effective potential for the system. Through effective potential techniques, we find that high density attractive and repulsive BECs exhibit new character with direct impact over the MI phenomenon. Results of intensive numerical investigations are presented and their convergence with the above semi analytical approach  is brought out.
\end{abstract}
\maketitle
\section{Introduction}
\label{sec1}

Solitons in a Bose-Einstein condensate (BEC) trapped in an optical lattice (OL) potential have attracted great attention recently. Generally, BEC in OL is an ideal test bed for condensed matter theory with non-linearity, which gives variety of phenomena such as gap solitons, localization, breathers, diffusion, vortices in lattice, etc... Such a rich dynamics that arises in BECs in OL can be attributed to the phase transition that takes place from superfluid to mott-insulator. In particular, the dispersion dynamics in OL can give rise to stable localized matter wave states in the form of gap solitons. This is represented by stationary solutions of the associated  Gross-Pitaevskii (GP) equation with the eigenvalue located at optically induced finite bandgap, in the repulsive condensate also~\cite{Marsh2006}.

In the ultracold regime, most of the results of experiments in BECs are reproduced and described by the theoretical model based on the nonlinear mean-field GP equation with two-body interaction~\cite{Dalfovo1999,Sabari2010} and the number density dependent three-body interaction~\cite{Abdullaev2001,Garcia2003}.
However, it should be pointed out that the impact of higher order interactions have not yet been considered~\cite{Dalfovo1999,PRE86,Gammal2000}. The  fact that  the BEC density increases in various experiments with strong compression of the trap,~\cite{Ping2009,Sabari2015} makes the introduction   of higher-order interaction  more realistic and inevitable in the description of the dynamics. In very recent findings, higher-order interactions challenged the three-body interactions over the MI of BECs~\cite{OlivierJOSAB2020}. Earlier studies bring about stabilization in BECs even with repulsive two body interaction with/without nonlinear management in which the role of higher order interactions has been made more crucial~\cite{Tam_stabPLA2019,Tam_stabJPB2018} 

The inclusion of the shape dependent higher-order interaction  at higher density through residual nonlinearity is justified as new phenomena  and may bring about  changes into the dynamics of BECs especially when immersed in traps like OL potentials. It can be  recalled  that the  recent  first real time solitary matter wave pulse was observed experimentally in Rb atoms by the process of modulational instability, where the condensate was kept in periodically varying optical lattice structure (optical waveguide).
Thus, in the present work, we examine the effects of higher-order interactions and how they impact the MI~\cite{Benjamin1967} of plane wave for a collective interactions including residual nonlinearity  of the  system consisting of BECs trapped in an OL potential driven by harmonic trap. To produce localized patterns such as solitons, breathers and fundamental vortices which form the  basis for energy transport mechanism in nonlinear physical systems, the investigation of the onset of  MI is inevitable.
Some of the recent investigations towards the  exploration of MI parametric domain by means of linear stability analysis from the perspective  of higher order interactions in recent years include  BECs with zero-nonlinearity in single component~\cite{IA_Bhat}, residual nonlinearity in two-component condensate~\cite{Tam_MIPRE}, Spin-Orbit coupled (SOC) BECs mixture~\cite{Panigrahi}, inter-component asymmetry interaction in quantum droplets and SOC-BECs in optical lattice~\cite{Mithun,DIS_SOC_MI}.

In the present investigation, through the time-dependent variational approach (TDVA)~\cite{Rapti2004a,Sabari2013,OlivierJOSAB2020} and numerical calculations, we study the MI in BECs through the solutions of GP equation with the effect of periodic potential. For this purpose, we perform the TDVA to propose not only the MI conditions but also the time-dependence of the perturbation parameters. 
The paper is structured as follows: In Section \ref{sec2}, we present the theoretical model that describes the condensates under our consideration. Section \ref{sec3} is devoted to the mathematical framework in which we derive the MI conditions of the system through the TDVA. Then, in Section \ref{sec4}, we discuss the onset of MI under the higher-order interaction and OL potential. In Section \ref{sec5}, we perform direct numerical simulation to check the validity of the MI conditions found by analytical methods. Finally, in Section \ref{sec6}, we summarize our results and present our conclusions.

\section{The model}
\label{sec2}

At ultra-low temperatures, BECs with two-, three- body and higher-order interactions can be described by the following GP equation~\cite{PRE86,Wamba2013}

\begin{align}
\mathrm{i}\hbar\frac{\partial \Psi(\mathbf{r},t)}{\partial t} =&\,\left(-\frac{\hbar^2}{2m}\nabla^2+ V_{trap} +g |\Psi(\mathbf{r},t)|^2 \right)\Psi(\mathbf{r},t)\nonumber\\ &\,+ \left(g_3 |\Psi(\mathbf{r},t)|^4+ \eta_0 \nabla^2 |\Psi(\mathbf{r},t)|^2 \right)\Psi(\mathbf{r},t),\,\,\, \label{Jac1}
\end{align}
where $\hbar$ is the reduced Planck's constant,, $m$ is the mass of the boson, $V_{trap}=\frac{m}{2}(\omega_\rho^2 \rho^2+\omega^2_x x^2)+V_{opt}(x)$ upon which we emphasize  that the OL potential is applied only in the longitudinal direction, i.e., $V_{opt}(x)=V_{m}\cos^2(kx+\theta)$, with $V_{\mathrm{m}}$ being the potential depth of OL. The parameter $\theta$ is an arbitrary phase and $k={2\pi}/{\lambda}$ is a wave number of the OL that can be controlled by varying the angle between two counter propagating laser beams whose interference creates the OL~\cite{Olivier2020}.  $\omega_x$ and $\omega_{\rho}$, respectively, are the longitudinal and radial frequencies of the external trap, and $\rho$ denotes the radial distance. The parameters $g$ and $g_3$ are the strengths of the two- and three-body interatomic interactions, respectively.
The role of $g$ depends on $a_s$ by $g=4\pi\hbar^2 a_s/m$.
The last term describes the shape-dependent confinement correction of the two-body collision potential. The parameter $\eta$ is the higher-order scattering coefficient which depends on both the \textit{s}-wave scattering length and the effective range for collisions~\cite{Zinner2009,Collin2007}. This parameter reads $\eta_0= g g_2$, where $g_2$ is defined by $g_2=a_s^2 /3-a_s r_e /2$, with $r_e$ being effective range.

The radial motion can be strongly confined by making the radial trapping frequency $\omega_{\rho}$ much larger than the axial frequency $\omega_x$. In this case , the condensate is cigar-shaped, and owing to that, one can take 
\begin{eqnarray}
\Psi(\mathbf{r},t) = \phi_0(\rho)\phi(x,t)
\end{eqnarray}
with
$\phi_0=\sqrt{1/(\pi
a^2_{\perp})}\exp\left(-\rho^2/(2a^2_{\perp})\right)$, $\rho=\sqrt{y^{2}+z^{2}}$ and $a_{\perp}=\sqrt{\hbar/m\omega_{\perp}}$, is the ground state of the radial equation,
\begin{eqnarray}
-\frac{\hbar^2}{2m}\nabla^2_\rho \phi_0+ \frac{m}{2}\omega^2_{\rho}
\rho^2 \phi_0 = \hbar \omega_{\rho}\phi_0.
\end{eqnarray}
Then, multiplying both sides of the GP equation (\ref{Jac1}) by $\phi^*_0$
and integrating over the transverse variable $\rho$, we can get the
following quasi-1D GP equation \cite{PRE86,OlivierJOSAB2020,Olivier2020}:
\begin{align}
\mathrm{i}\hbar\frac{\partial \phi(x,t)}{\partial t} &=\left(-\frac{\hbar^2}{2m}\frac{\partial^2}{\partial x^2}+ \frac{m}{2} \omega_x^2 x^2+V_{m} cos^2(\kappa x)\right)\phi(x,t)\nonumber\\ &\,+ \left(g\prime |\phi|^2+ g_3\prime |\phi|^4+ \eta_0\prime \frac{\partial^2 |\phi|^2}{\partial x^2}  \right)\phi(x,t), \label{Jac4}
\end{align}
where, $g\prime=g/(2\pi a_{\perp}^2)$, $g_3\prime=g_3/(3\pi^2 a_{\perp}^4)$ and $\eta_0\prime=\eta_0/(2\pi a_{\perp}^2)$. 
It is more convenient to use the above Eq. (\ref{Jac4}) into a dimensionless form. For this purpose, we make the transformation of variables as $\tilde{t}=t\mu$, $\tilde{x}=x\kappa$, $\psi=\phi\sqrt{2a_{so}\omega_{\perp}/\mu}$, where $\mu=E_R/\hbar$ with $E_R=\hbar^2\kappa^2/2m$. In the case of deep optical potential wells combined to a weak confinement, the harmonic trapping potential can be neglected compared to the OL potential~\cite{Sabari2013}. In this case, we can get the following normalized 1D GP equation:

\begin{align}
\mathrm{i}\frac{\partial \psi(x,t)}{\partial t} =&\,\left(-\frac{\partial^2}{\partial x^2}+V_{s} cos^2( x)+g |\psi(x,t)|^2\right)\psi(x,t)\nonumber\\ &\,+ \left( \chi |\psi(x,t)|^4+ \eta \frac{\partial^2 |\psi(x,t)|^2}{\partial x^2}  \right)\psi(x,t), \label{Jac5}
\end{align}
$V_s = V_m/E_R$, $g= a_s/a_{so}$, $\chi=g_3\prime a_{so}^2\omega_{\perp}^2\kappa^4/E_R$, and $\eta = g g_2 \kappa^2$, where $a_{so}$ is the constant scattering length.

\section{Modulational instability}
\label{sec3}
Making use of  the variational approach in ref.~\cite{Sabari2013,OlivierJOSAB2020}, we write the lagrangian density as
\begin{eqnarray}
\mathcal{L} = &\frac{\mathrm{i}}{2}\left(\frac{\partial \psi}{\partial
t}\psi^*-\frac{\partial \psi^*}{\partial
t} \psi\right)-\left|\frac{\partial \psi}{\partial z}\right|^2 -V_s\cos^2(x)|\psi|^2\nonumber\\ &\,-\frac{1}{2}g|\psi|^4 -\frac{1}{3}\chi|\psi|^6-\frac{1}{2}\eta\frac{\partial^2|\psi|^2}{\partial z^2}|\psi|^2. \label{Jac7}
\end{eqnarray}
with the MI-motivated wavefunction of the form
\begin{eqnarray}
\psi(x,t)= (A_0 + \delta) \exp[\mathrm{i}(kx-(k^2&+g A^2_0+\eta A^2_0
\nonumber\\ &+\chi \,A^4_0)t)],\label{Jac8}
\end{eqnarray}
where $\delta=a_1(t)\exp[\mathrm{i}(q x+b_1(t))] + a_2(t)\exp[\mathrm{i}(-q x+b_2(t))]$.
Using this ansatz in the procedure used in Refs.~\cite{Rapti2004a}, the following equations are obtained (derivations given in Appendix A)
\begin{eqnarray}
\frac{\partial a}{\partial t} = A_0^2 a\sin(b)[q^2 \eta -g-\chi(2A_0^2 +6 a^2)],\label{Jac19}
\end{eqnarray}
\begin{eqnarray}
\frac{\partial b}{\partial t} =-2q^2-V_s+2A_0^2 q^2 \eta-4A_0^4 \chi  +8q^2 \eta a^2 \nonumber\\ - \chi\, a^2\,(36\, A_0^2+20 \,a^2) -2\,g\left(A_0^2+3\,a^2\right) \nonumber\\ +  2 A_0^2 \cos(b)[q^2 \eta-g-\chi(2A_0^2+12 a^2)],\label{Jac20}
\end{eqnarray}
with the effective potential given by

\begin{eqnarray}
V_{\mathrm{eff}}&= 2A^2\,V_s \Big[q^2+A_0^2\,(\,g-q^2\,\eta+2A_0^2\,\chi\,)+\frac{V_s}{4}\Big]  \nonumber\\ &+2A^2 \Big[q^2\,(q^2+2\,g\,A_0^2+4A_0^4 \,\chi-2q^2\,\eta\,A_0^2 \,)\Big]  \nonumber\\ &+ 3A^3\Big[g(\,\frac{1}{2}V_s +q^2+g\,A_0^2)  +\chi A_0^2(3V_s+6 q^2\nonumber\\ & +\,4\, g\,A_0^2\,)+\,4 \,A_0^4\,)\Big]-\,\eta\,A^3\,\Big[4\,q^4-\,4\,q^4\,A_0^2\,\eta \nonumber\\ & +2q^2V_s +7A_0^2q^2(g +2A_0^2)\Big] + A^4\Big[\frac{\chi}{3}(5V_s +10q^2  \nonumber\\ & +\frac{175}{2}\,A_0^4\,\chi+\frac{101}{2}\,A_0^2\,g)+\frac{9}{8}\,g^2\Big] +\frac{25}{18}\,\chi^2\,A^6\nonumber\\ &-\eta\,A^4\,\Big[-2\,q^4\, \eta +\frac{64}{3}\,A_0^2\,q^2\,\chi+\,3\,q^2\,g\,)\,\Big]\nonumber\\ &+A^5\chi \Big[\frac{5}{2}g+15 A_0^2\chi-\frac{10}{3}q^2\eta\Big]. \label{Jac25}
\end{eqnarray}
Common derivation of stability analysis sets the time-dependent condition for the modulationally unstable waves with  wavenumber $q$ evolving in BECs trapped in an OL potential as:
\begin{eqnarray}
V_s^2&+4q^2(V_s+q^2+2gA_0^2+4\chi A_0^4-2\eta q^2 A_0^2) \nonumber \\ & +4A_0^2V_s(g+2A_0^2 \chi - q^2 \eta) <0. \label{Jac26}
\end{eqnarray} 
\color{black}
\begin{figure}[hb!]
 \begin{center}
 \includegraphics[width=8.5cm,height=4.5cm]{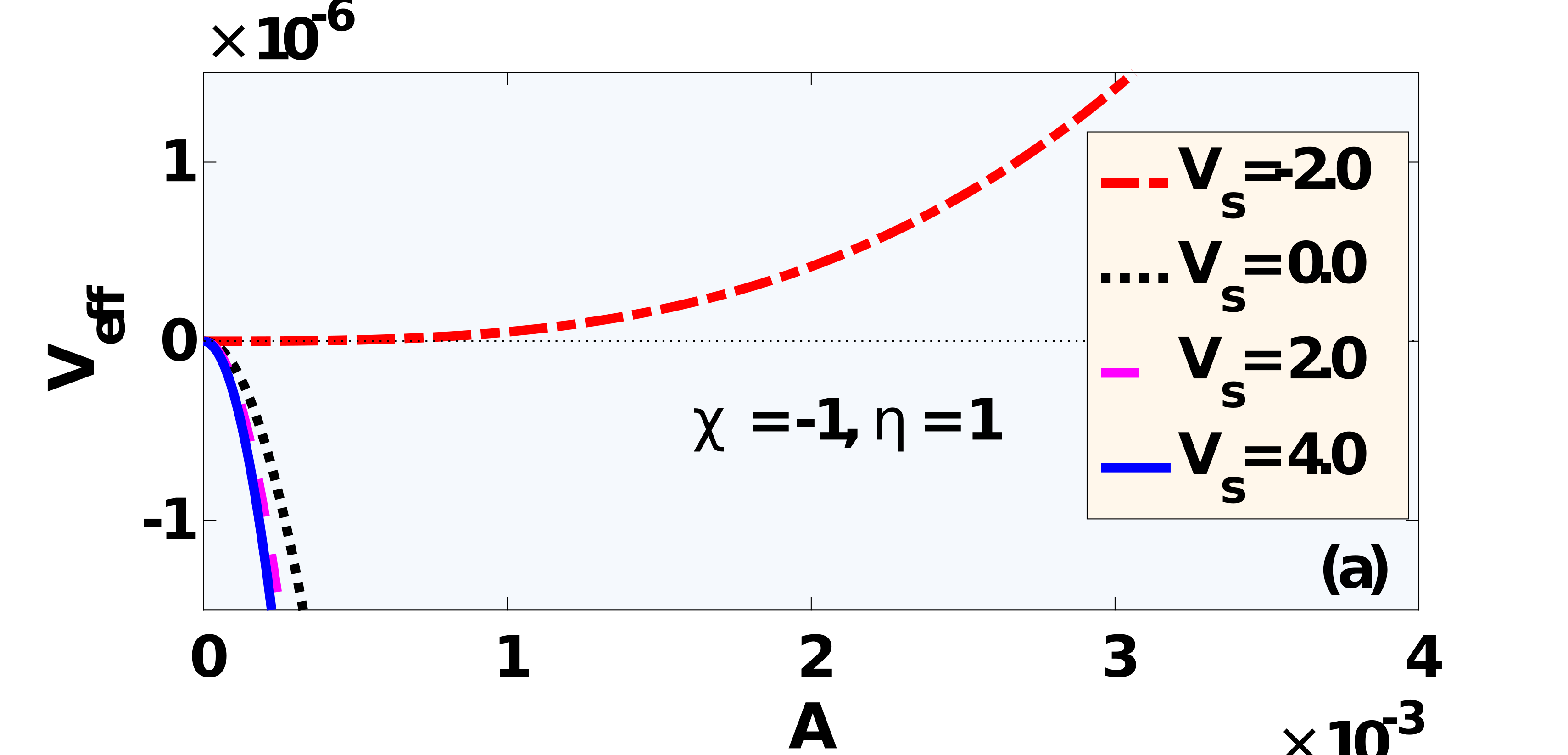}
 \includegraphics[width=8.5cm,height=4.5cm]{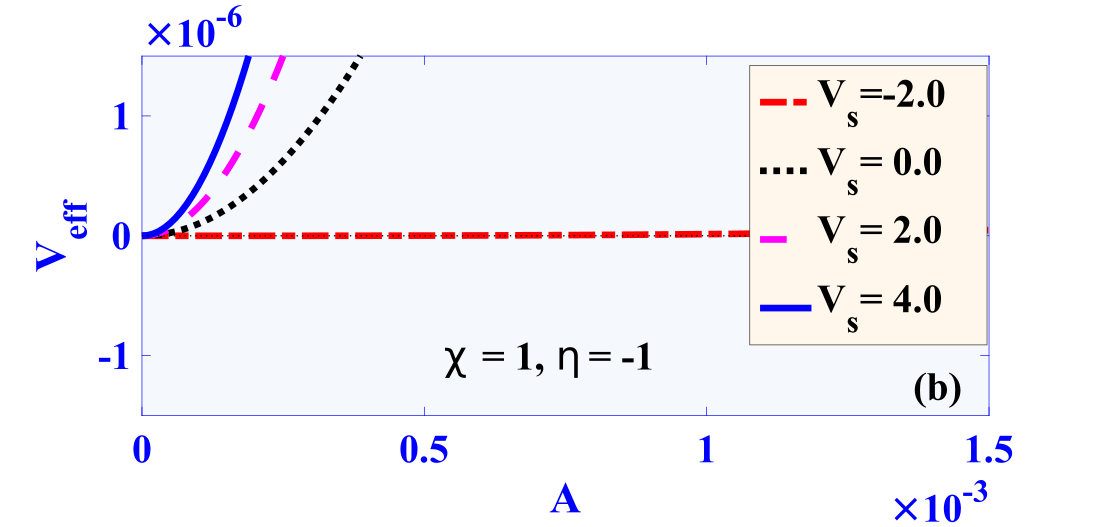}
  \end{center}
 \caption{Effective potential $V_{eff}$ Vs perturbed amplitude $A$ for attractive two body interaction $g=-1$. (a) Negative three-body interaction $\chi=-1$ and positive higher order interaction $\eta=1$. (b) positive three-body interaction $\chi=1$ and negative higher-order interaction $\eta=-1$. All quantities plotted are dimensionless.}
 \label{f1}
 \end{figure}

\begin{figure}[hb!]
 \begin{center}
 \includegraphics[width=8.5cm,height=4.5cm]{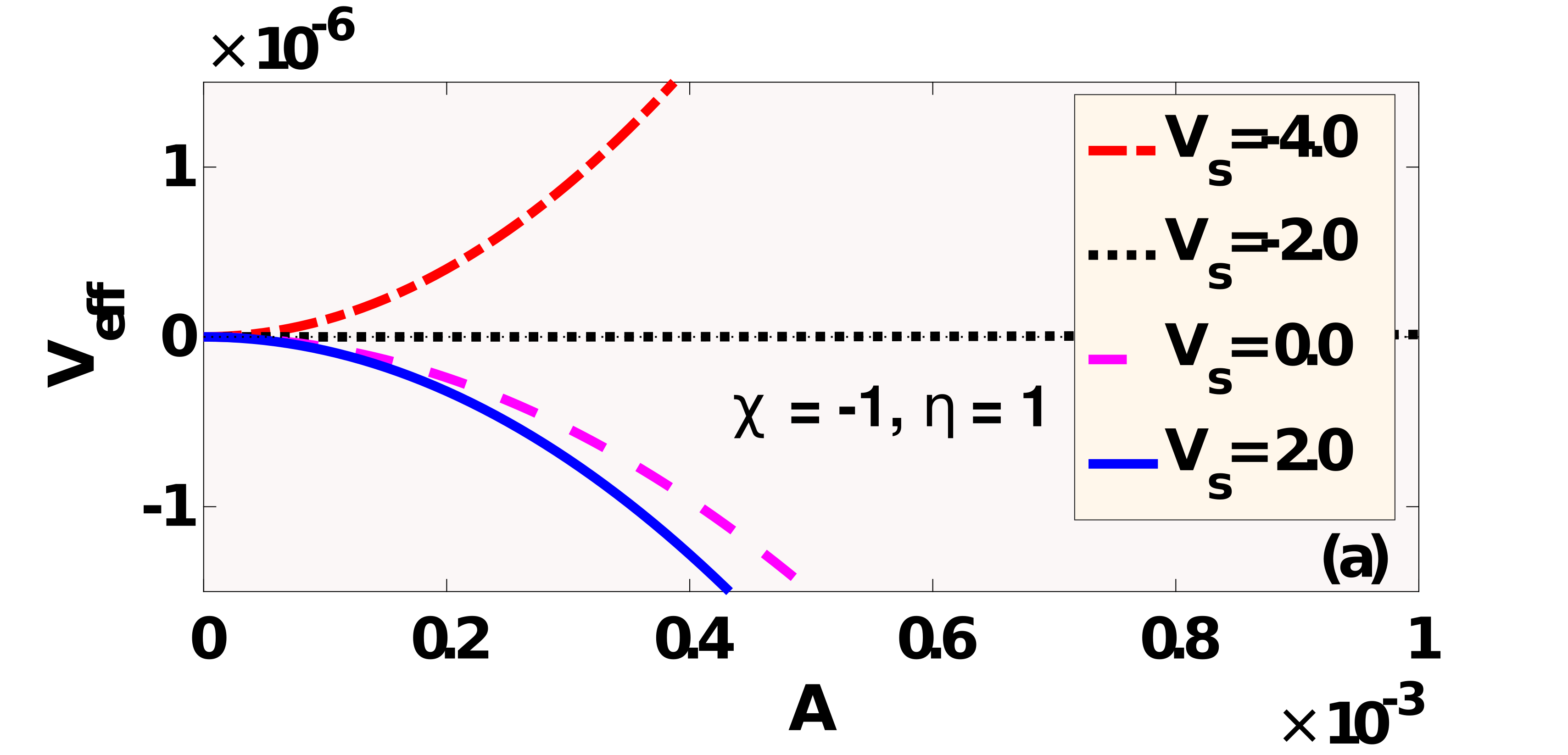}
 \includegraphics[width=8.5cm,height=4.5cm]{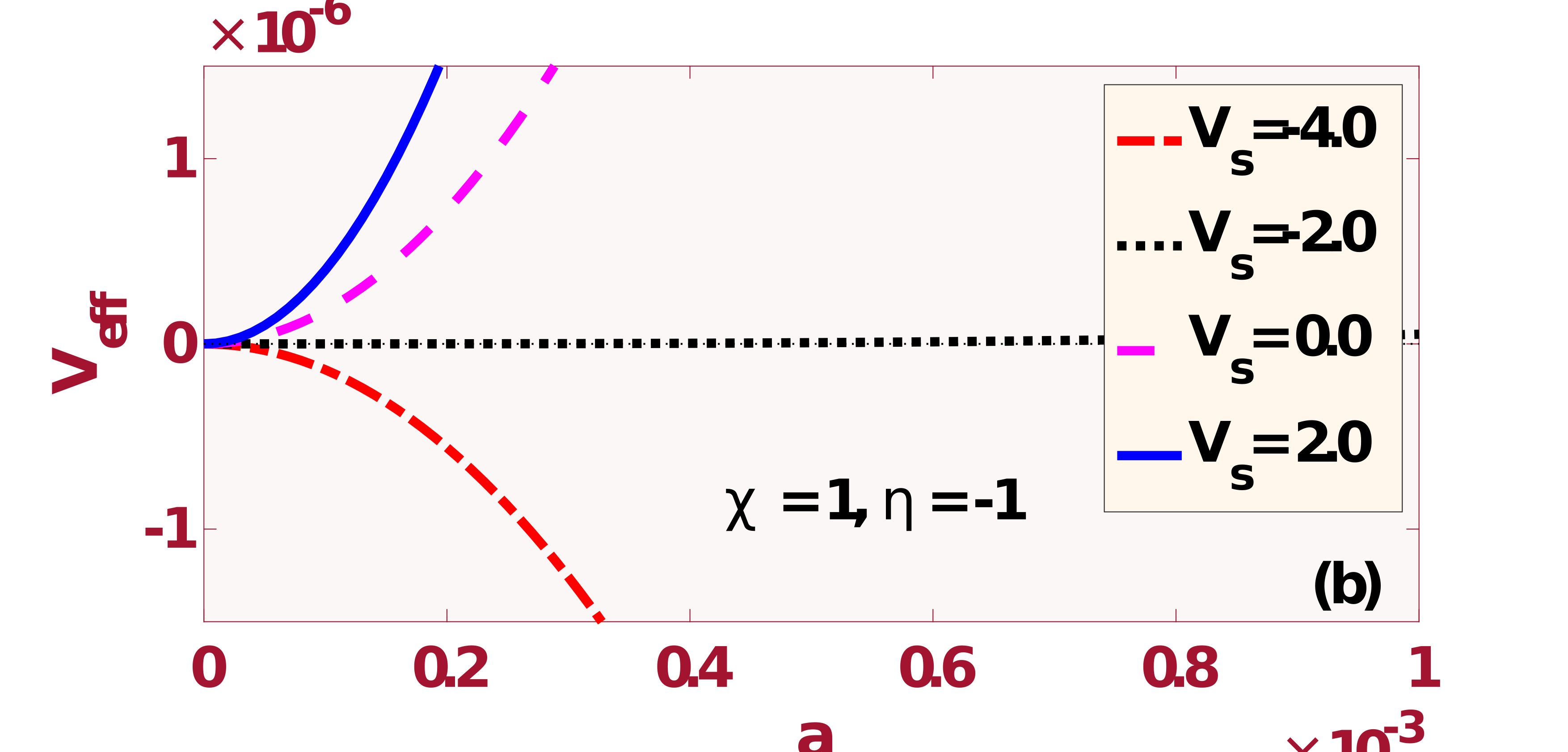}
  \end{center}
 \caption{Effective potential $V_{eff}$ Vs perturbed amplitude $A$ for repulsive two body interaction $g=1$ . (a) Negative three-body interaction $\chi=-1$ and positive higher order interaction $\eta=1$. (b) positive three-body interaction $\chi=1$ and negative higher-order interaction $\eta=-1$. All quantities plotted are dimensionless.}
 \label{f2}
 \end{figure}

\section{Onset of MI under Higher-order interactions and  OL potential}
\label{sec4}

Next, through the effective potential, we can analyze the role of the three-body interaction on the stability of the BECs along with both two-body and higher-order interactions. 
One can determine the stability of the dynamics from the assessment of the curvature of effective potential at $A = 0$. The dynamics is stable if the potential is convex, i.e. the second derivative $\frac{\partial^{2}
V_{\mathrm{eff}}}{\partial A^2}|_{A=0}>0$. Otherwise ,the dynamics is unstable if the potential turns out to be concave, i.e. the second derivative $\frac{\partial^{2}
V_{\mathrm{eff}}}{\partial A^2}|_{A=0}<0$.

In FIG.\ref{f1}, the effective potential curves are shown for the attractive two-body condensate. Panel (a) corresponds to the  combination of negative three-body interaction ($\chi=-1$) and  positive higher order interaction ($\eta=1$)  for four different strengths of OL range chosen as $V_s\in[-2,4]$. The wavenumber $q$ is modulationally stable for $V_s=-2$ and  modulationally unstable for $V_s=0,2,4$.
In panel(b), for repulsive three-body ($\chi=1$) and attractive higher-order interactions ($\eta=-1$) with the same range of optical potential $V_s\in[-2,4]$, the wavenumber gives stable modes for optical strength $V_s= 0,2,4$ and critically stable mode for $V_s= -2$. One can easily observe that for any change of sign of interaction, the stability of the system shows some kind of anti-symmetry nature in MI scenario. The physical insight has been observed from effective potential curve when we exchange the sign of higher-order interaction for attractive two-body condensates. The stability condition is also controlled  by the strength of optical lattice energy reinforced with short range interactions and accordingly,  the instability gets either enhanced or  suppressed. 

In FIG.\ref{f2}, the effective potential curves are shown for the repulsive two-body condensates. Panel (a) corresponds to the  combination of negative three-body interaction ($\chi=-1$) and  positive higher order interaction ($\eta=1$)  for four different strengths of OL range chosen as $V_s\in[-4,2]$. The wavenumber $q$ is modulationally stable for $V_s=-4$, critically stable for $V_s= -2$, and unstable for $V_s=0,2$.
In panel(b), for repulsive three-body ($\chi=1$) and attractive higher-order interactions ($\eta=-1$) with the same range of optical potential $V_s\in[-4,2]$, the wavenumber gives stable modes for optical strength $V_s= 0,2$, critically stable mode for $V_s= -2$, and unstable mode for $V_s= -4$.

Thus, the repulsive two-body condensate also shows the anti-symmetric  MI behavior with respect to the reversal of the sign of interaction strength similar to  attractive condensates which means that energy excitation over the initial perturbation has exchanged symmetry with respect to three-body and higher-order interactions controlled by the strength of  optical potential. Moreover, if we turn off, the higher-order interaction also, we  find  the stability condition as in~\cite{Sabari2015}. The removal of  three-body interaction in our model results in the instability condition found  in ~\cite{mitheo2,Theocharis2003}. The results of MI in a harmonic trap can also be obtained as a special case of our model when we remove   optical potential  ~\cite{Wamba2013}.
\begin{figure}[!t]
\begin{center}
\includegraphics[width=9.5cm,height=5cm]{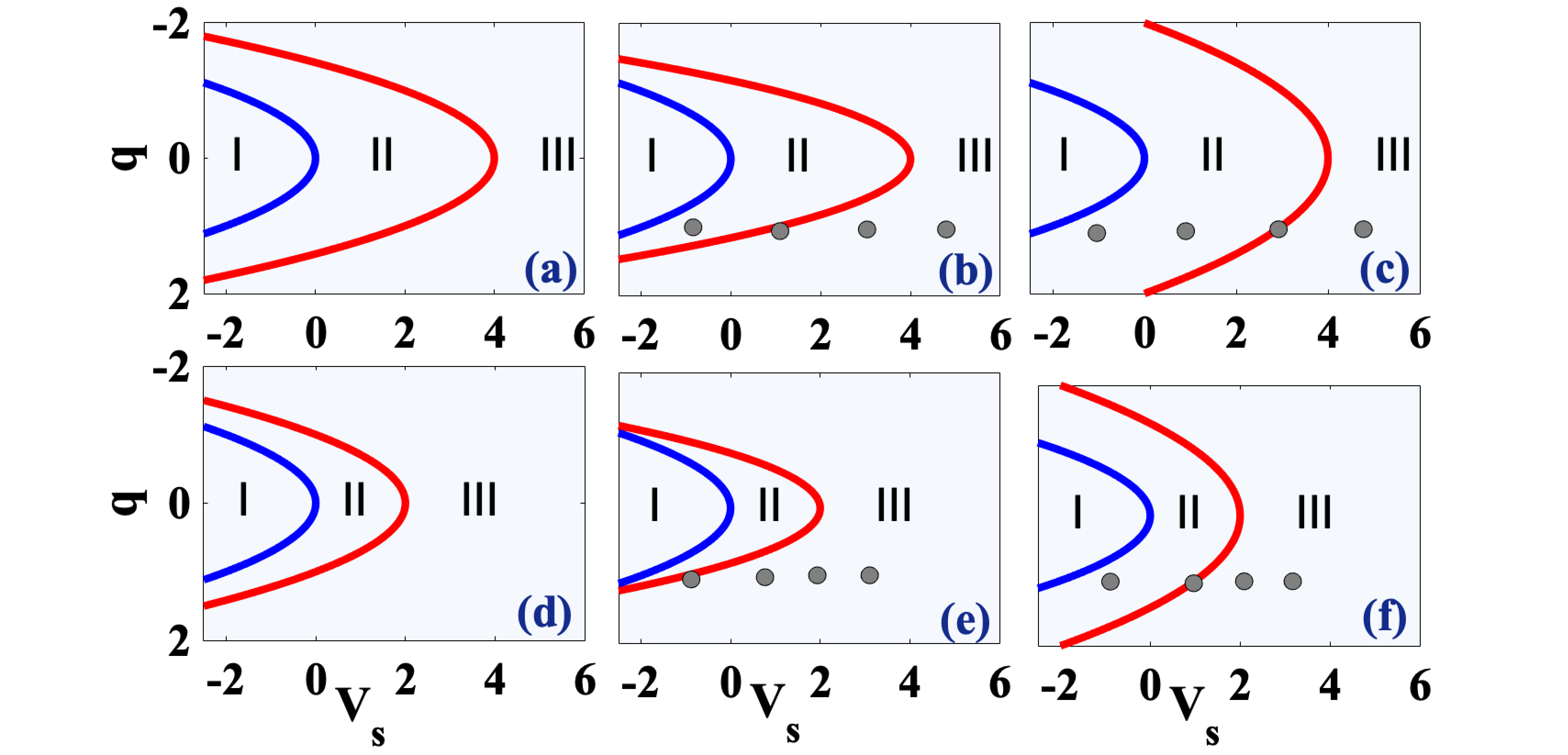}
\end{center}
\caption{ The stability/instability domain in the ($V_s,q$) plane for attractive two-body interaction i.e., $g<0$. From left to right ([a,b,c] and [d,e,f]), $g_2=0$, $g_2=0.25$ and $g_2=-0.25$, respectively. The upper and lower rows correspond to $\chi=0$ and $\chi=0.25$, respectively. All quantities plotted are dimensionless.}
\label{f3}
\end{figure}
FIG.3 shows the modulationally stable and unstable domains in the ($V_s,q$) plane for attractive two-body interaction ($g<0$). The top row in  FIG.3 shows absence of three body interaction ($\chi=0$) for   $g_2=0$, $g_2=0.25$ and $g_2=-0.25$ from left to right respectively. Here, the three different regions made by the parabolas which are obtained through  Eqs.(\ref{Jac26}) represent stability domains for finite range of  modulational wave number. The region (II) which is in between two parabolas indicates unstable domain while the region (I) surrounded by first parabola and also the region (III) behind the second parabola represent the stable domains. When the higher-order interaction is turned off ($g_2=0$), the  modulational unstable region is quite large. For positive higher order interactions ($g_2=0.25$), the wave number shrinks in the tail of two parabolas which implies it suppresses instability for appropriate negative values of optical potential strength and the location of vertex of two parabolas is similar as in the case of absence of higher-order interaction. But for  negative higher order interaction ($g_2=-0.25$), the tails of the parabolas  expand so that the area of the unstable region increases, which causes the enhancement of MI on modulated wave number ($q$). Hence, the negative higher-order interaction  with attractive two-body interaction supports  system instability.
\begin{figure}[!t]
\begin{center}
\includegraphics[width=9.5cm,height=5cm]{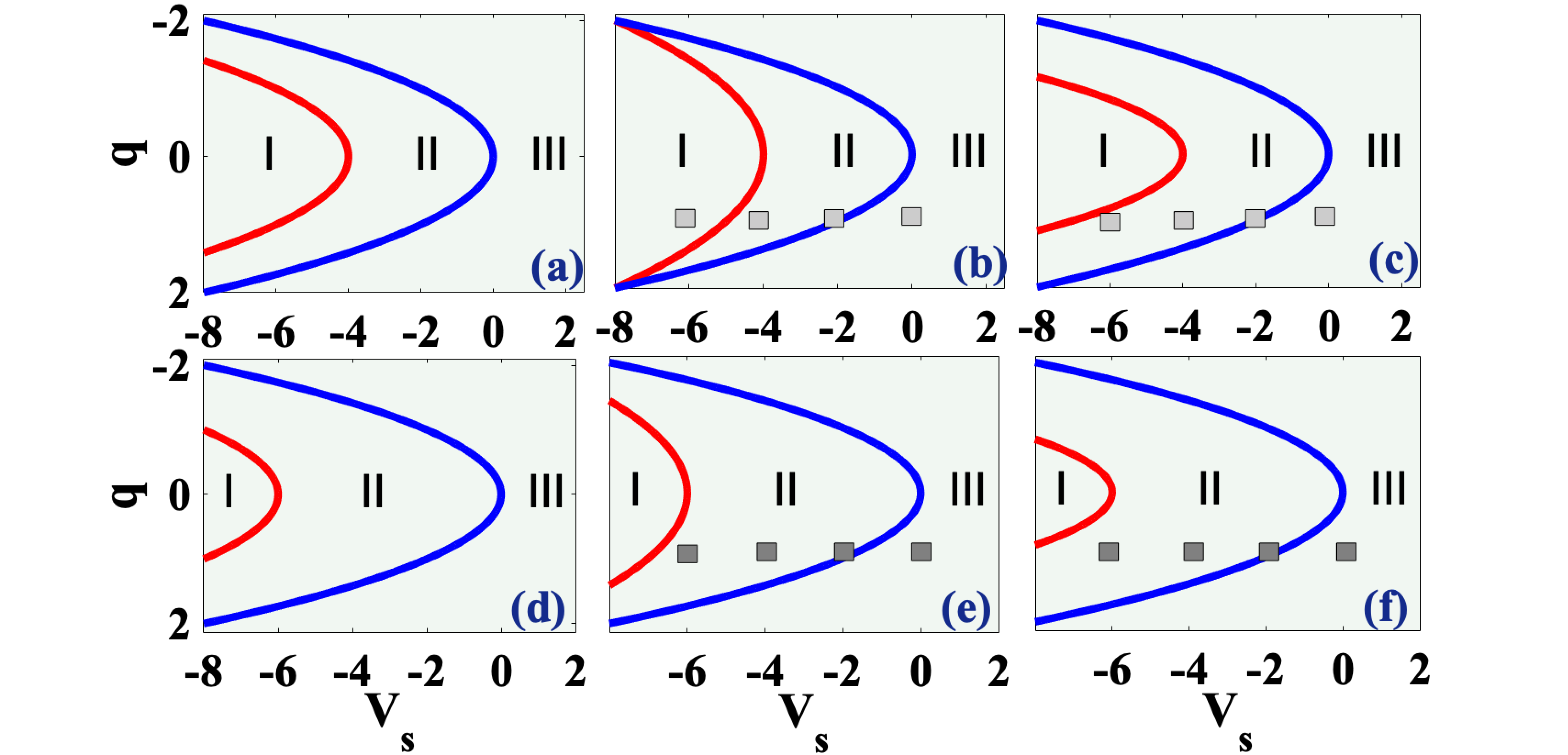}
\end{center}
\caption{ The stability/instability domain in the ($V_s,q$) plane for repulsive two-body interaction i.e., $g>0$. From left to right ([a,b,c] and [d,e,f]), $g_2=0$, $g_2=0.25$ and $g_2=-0.25$, respectively. The upper and lower rows correspond to $\chi=0$ and $\chi=0.25$, respectively. All quantities plotted are dimensionless.}
\label{f4}
\end{figure}
Now when we turn on the three-body interaction as positive ($\chi=0.25$), the ($V_s, q$) domain is shown in bottom of FIG.3. The remaining parameters are the same as what we used in upper panel. In this case, similar to  the upper panel, the parabolas have three regions (I,II,III). In this case, the vertex of second parabola shifts towards zero of optical potential strength for all the higher-order interaction values ($g_2=0,0.25,-0.25$). The area of region (II) in the bottom row, which represents the unstable domain has reduced . The inclusion of positive three-body interaction with higher-order interaction of attractive condensate suppresses the system instability.
In Fig.4, which is also obtained from Eqs.(\ref{Jac26}), the ($V_s, q$) domain has been shown for repulsive two-body interaction ($g>0$) condensate in the absence of three-body interaction (top row, $\chi=0$) and positive three-body interaction (bottom row, $\chi=0.25$). In  both top and bottom rows, from left to right, higher-order interaction strength is $g_2=0$, $g_2=0.25$ and $g_2=-0.25$ respectively.  In the absence of three-body interaction ($\chi=0$) for all  values of higher order interaction, the domain has two modulationally stable regions (I, III) and one unstable region (II). For  $g_2=0$ and $g_2=-0.25$, they have similar kind of MI domain as shown in between the vertex of two parabolas. But, the domain for $g_2=-0.25$ has  expanded more at the tails of the parabolas so that the MI has enhanced at this value of potential strength. The positive higher-order interaction ($g_2=0.25$) suppresses the MI near the tails of the two parabolas where the shrink has been more pronounced, but between the vertex, it supports instability of the system.
\begin{figure}[!t]
\begin{center}
\includegraphics[width=9.5cm,height=4.55cm]{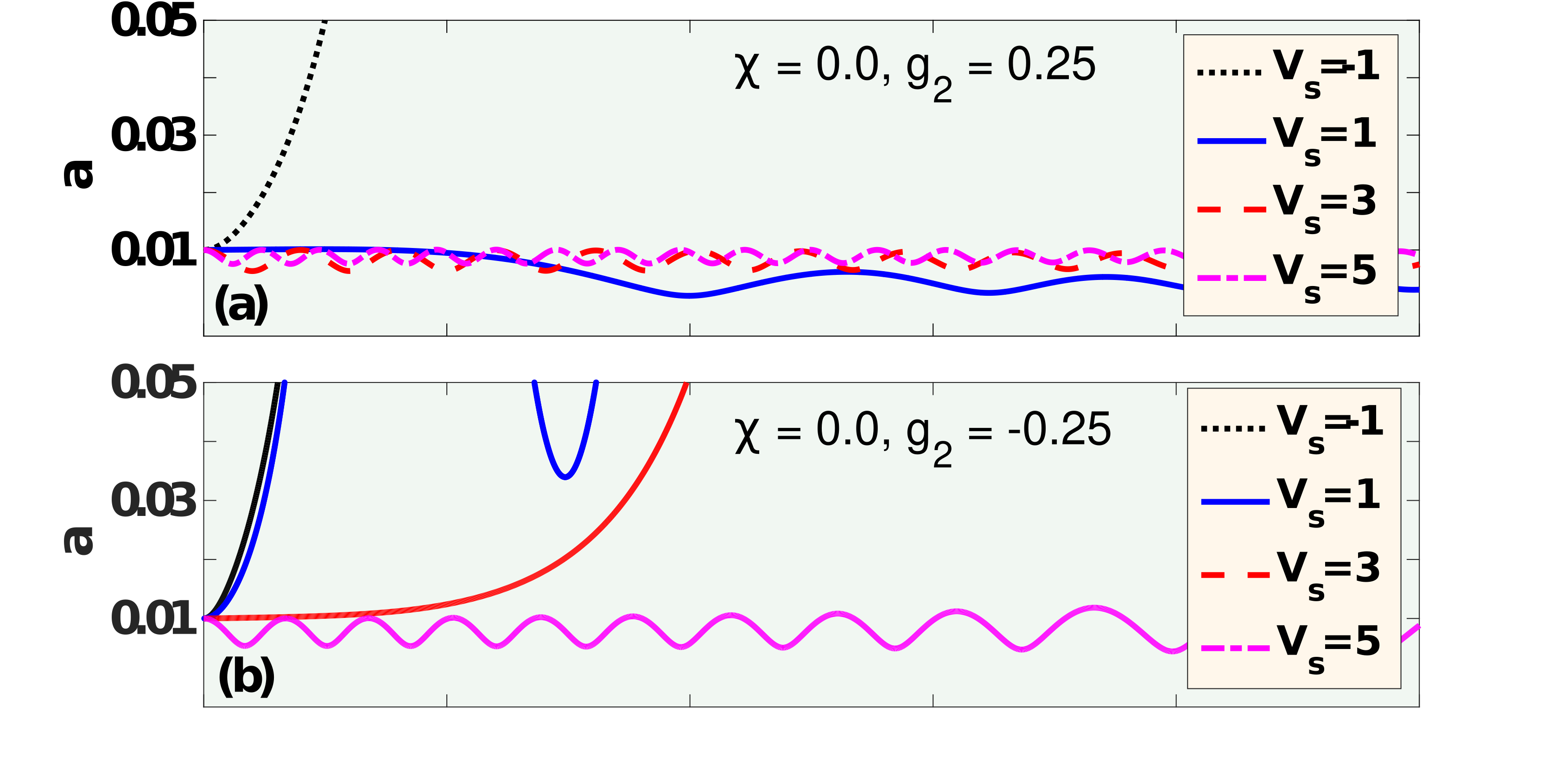}
\includegraphics[width=9.5cm,height=4.5cm]{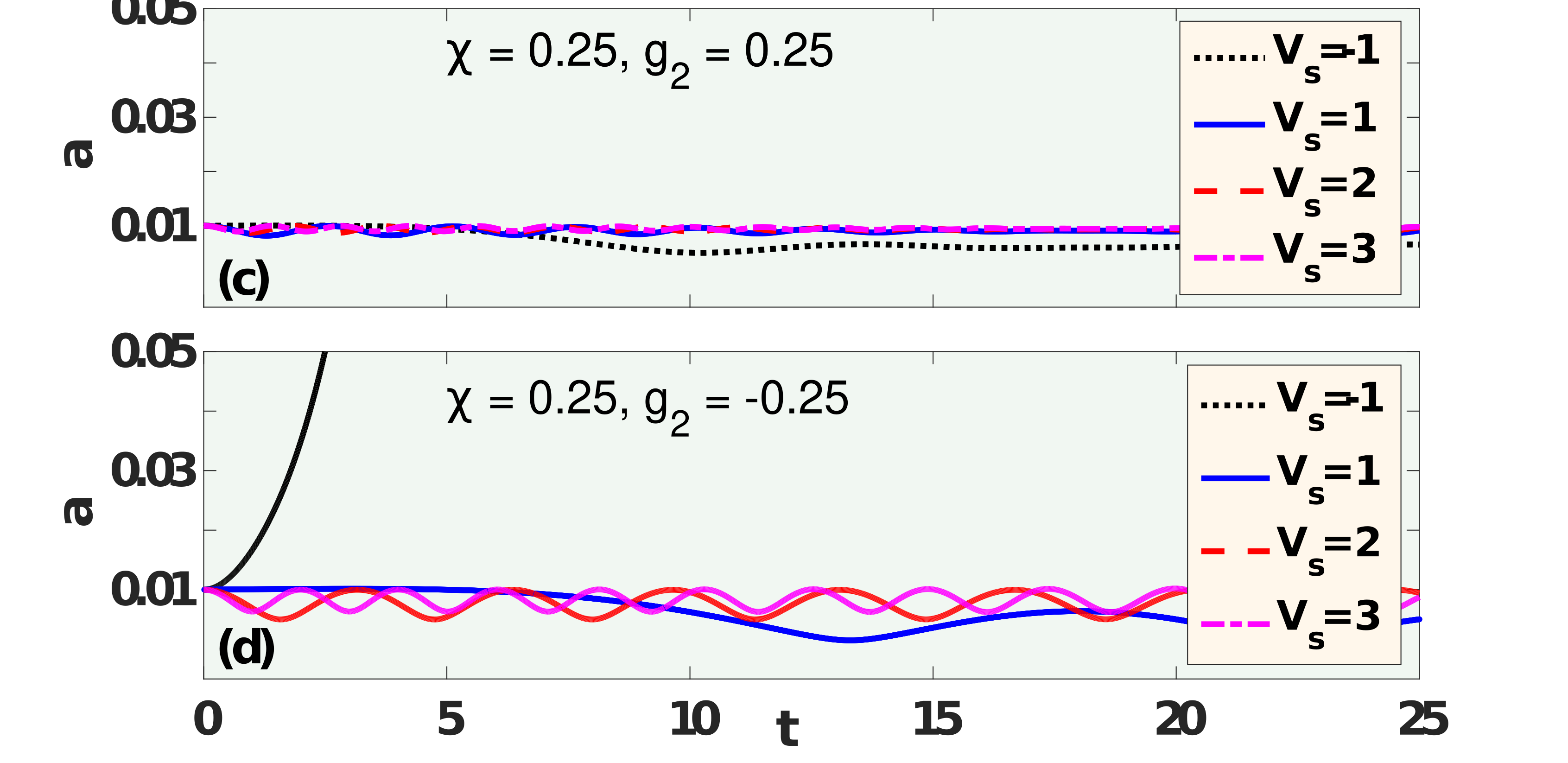}
\end{center}
\caption{ Response of variational parameter ($a$) with respect to time ($t$) for attractive two-body interaction $g<0$. (a) and (b) for absence of three body interaction($\chi=0$)  with $g_2=0.25$ and $g_2=-0.25$ respectively. (c) and (d) for presence of three-body interaction ($\chi=0.25$) with $g_2=0.25$ and $g_2=-0.25$ respectively. All quantities plotted are dimensionless.}
\label{f5}
\end{figure}
In the bottom row of FIG.4, the three-body interaction has been made positive ($\chi=0.25$) with the   higher-order interaction strengths  ($g_2=0$, $g_2=0.25$ and $g_2=-0.25$) being the  same from left to right as in the case of absence of three-body interaction ($\chi=0$). The combined impact of three-body interaction and higher-order interaction along with repulsive two-body interaction  supports  modulationally unstable regimes for the negative optical potential strength. One can observe from the figure that  area of region II has been enhanced which corresponds to the negative domain of optical potential.  Here, we observe that the inclusion of the three-body interaction with higher order interaction in the repulsive condensate can enhance the system instability. In general, this does not support  instability dynamics in simple non-linear frame without trap (i.e. NLS equation)
%
\begin{figure}[!t]
\begin{center}
\includegraphics[width=9.5cm,height=4.5cm]{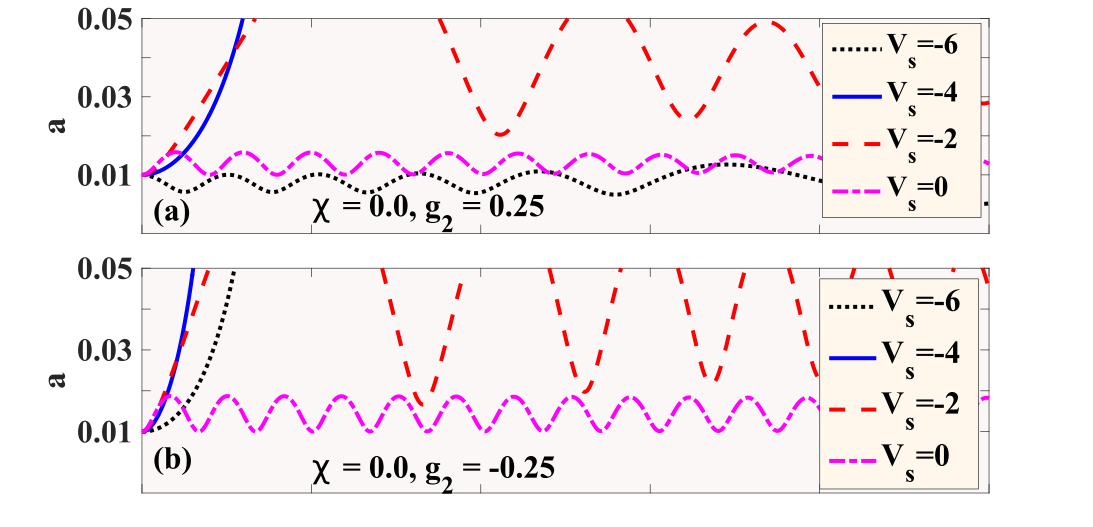}
\includegraphics[width=9.5cm,height=4.5cm]{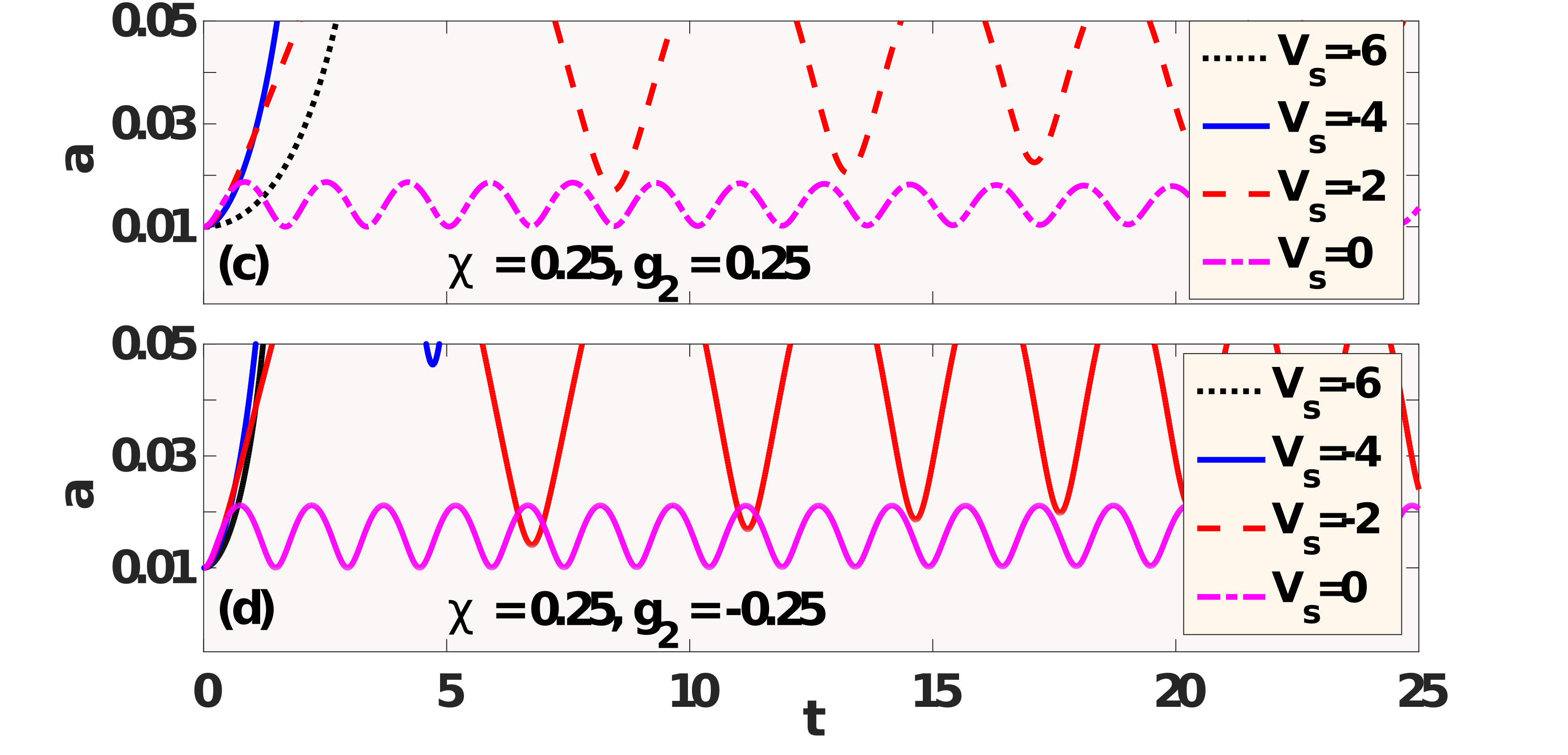}
\end{center}
\caption{Response of variational parameter ($a$) with respect to time ($t$) for repulsive two-body interaction $g>0$. (a) and (b) for absence of three-body interaction ($\chi=0$) with $g_2=0.25$ and $g_2=-0.25$ respectively . (c) and (d) for presence of three-body interaction ($\chi=0.25$) with $g_2=0.25$ and $g_2=-0.25$ respectively. All quantities plotted are dimensionless.}
\label{f6}
\end{figure}
Besides this, the instability conditions in ($V_s, q$) domain  from FIG.3 and FIG.4 can also be  justified by checking the time evolution equations of variational parameters $a(t)$ and $b(t)$ described by the ODEs in  Eqs.(7)-(8). The variational parameter $a(t)$ preserves its initial oscillatory profile for stable dynamics because the interplay doesn't happen between  non-linearity and  linear dispersion. If the interplay takes place in the system, the dynamics becomes unstable. In that case, the variational parameters  exponentially grow with time. Then, we numerically solve the set of equations given by eqs. (7)-(8) by means of a fourth order Runge-Kutta scheme. The initial condition values used here are $a(0)=0.01$ and $b(0)=0$. In FIG.5, we have shown the dynamics of the system in the presence of attractive two-body interaction. Panels (a-b) and (c-d) correspond to   the absence ($\chi=0$) and presence ($\chi=0.25$) of three-body interaction, respectively. Also, panels (a, c) and (b, d) pertain to  the presence of attractive ($\eta<0$) and repulsive ($\eta>0$) higher-order interaction, respectively. In each panel, we have shown the dynamics of the BECs for four different values of the optical potential which is chosen from FIG.3(b-c and e-f). In FIG.5(a), the initial perturbed amplitude  oscillates with time for optical potential strengths $V_s=1,3,5$ for $g_2=0.25$. Hence,it is obvious that the system is able to sustain  its stability for this set of parameters. It shows exponential growth mode for $V_s=-1$ which means that  excitation takes place and the system becomes unstable. 
In FIG.5(b) the dynamics shows unstable modes for optical potential strengths $V_s=-1$ and $1$ while stable modes exist up to $5$ time units  for $V_s=3$. The OL potential     for $V_s=5$ with $g_2=-0.25$ exhibits stable mode. The impact of the inclusion of three-body along with higher-order interactions is shown in FIG.5(c) and FIG.5(d) for the value of $g_2=0.25$ and $g_2=-0.25$ with OL potential strengths in the range $V_s\in[-1,3]$. The inclusion of three-body interaction in both cases ($g_2=0.25,-0.25$) enables the   initial amplitude  to  oscillate with time   against perturbation. The effects of higher-order interactions altogether suppress the instability dynamics in attractive condensates. 

\begin{figure}[!t]
\begin{center}
\includegraphics[width=4cm,height=8cm]{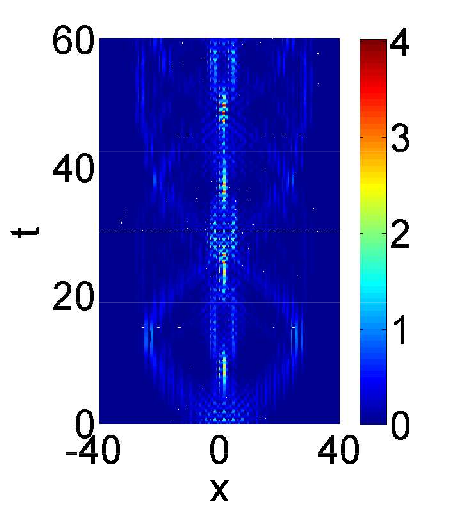}
\includegraphics[width=4cm,height=8cm]{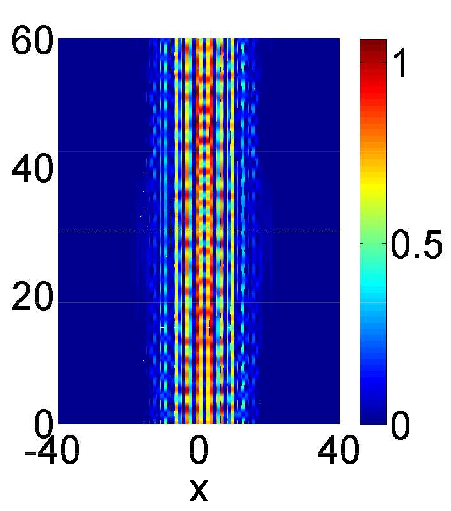}
\end{center}
\caption{The dynamics of the condensate wave for $g<0$ with $g_2=-0.25$ and $\chi=0.25$. The 3D space-time evolution of the square amplitude of the wave, $|\phi(x,t)|^2$, for 2 modes chosen in FIG.3(a). (Left panel) $V_s=0$ for unstable region and (right panel) $V_s=2$ for stable region. All quantities plotted are dimensionless.} 
\label{f7}
\end{figure}

The dynamics of the BECs with repulsive two-body interaction has been shown in FIG.6 (a) and (b) in the  absence of  three-body interaction along with $g_2=0.25$ and $g_2=-0.25$ respectively, while (c) and (d) correspond to the presence of three-body interaction along with $g_2=0.25$ and $g_2=-0.25$ respectively. In FIG.6 (a-d), we have shown the dynamics of the BECs for four different values of the optical potential ($V_s\in[-6,0]$) which is chosen from FIG.4(b-c and e-f). 
In FIG.6(a) and (b), the results are shown for the absence of three-body interaction along with $g_2=0.25$ and $-0.25$, respectively. In panel(a), the dynamics shows stability for $V_s=-6,2$  and instability for $V_s=-2,-4$. In panel(b), the dynamics shows stability for $V_s=0$ and instability for $V_s=-2,-4,-6$. 
In FIG. 6(c) and (d), the results are shown for the inclusion of three-body interaction with $g_2=0.25$ and $-0.25$, respectively. In both panels, the dynamics shows stability for $V_s=0$  and instability for increasing negative strength of optical potential $V_s=-2,-4,-6$. From all above results, one can observe that the  instability dynamics can be controlled by sign and strength of optical potential reinforced with higher-order interaction in high density condensates. This is well matched with analytical results from ($V_s, q$) domain (FIG.4 and FIG.5). \color{black}

\section{Numerical Results}
\label{sec5}
\begin{figure}[!b]
\begin{center}
\includegraphics[width=4cm,height=8cm]{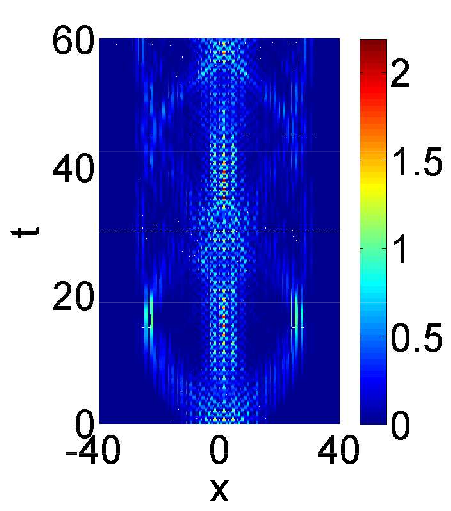}
\includegraphics[width=4cm,height=8cm]{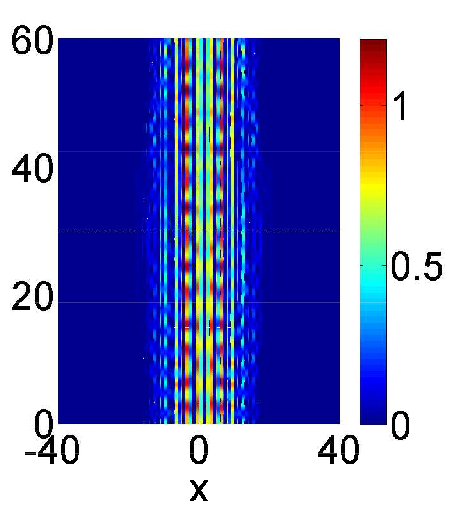}
\end{center}
\caption{The dynamics of the condensate wave for $g>0$ with $g_2=0.25$ and $\chi=0.25$. The 3D space-time evolution of the square amplitude of the wave for 2 modes chosen in FIG.4(e). (Left panel) $V_s=-4$ for unstable region and (right panel) $V_s=0$ for stable region. All quantities plotted are dimensionless}.
\label{f8}
\end{figure}
 The above analytical results are derived from  linearization of the unperturbed carrier wave. The validity of such analysis is limited to small amplitudes of perturbation compared with the carrier wave. So, analytical analysis of MI needs confirmation through direct numerical simulations of the modified GP Eq.(\ref{Jac4}). In the present study, we consider a limiting case of   the following equation, where the condensate is initially prepared in OL and embedded in harmonic magnetic field~\cite{mitheo2,Theocharis2003,Sabari2015}: $\phi(x,t)=\phi_{\mathrm{TF}}[\phi_{0}+\varepsilon \cos(q x)]$. Here, we use the Thomas-Fermi approximation considering $\phi_{\mathrm{TF}}=\sqrt{\mathrm{max}[0,\mu-V(x)]}$ as the background wave function. The numerical values are chosen as $\mu=1.0$, $\phi_0 \equiv A_0=1.0$, $ \alpha = 0.004$, $q=1.0$, $\varepsilon=0.001$ for both attractive and repulsive condensates during numerical integrations. The harmonic potential is only active at $t=0$, when the optical potential is turned off. However, the condensate is prepared at $t<0$ in an optically embedded harmonic magnetic field, where the initial wave function  affects the propagation without influencing its modulational instability or stability.
In FIG.7, we show the dynamics of 3D space-time evolution of the square amplitude of the wave, $|\phi(x,t)|^2$, for attractive condensate in OL embedded in harmonic potential with fixed values of higher-order non-linearity $g_2=-0.25$ and three-body interaction $\chi=0.25$. In order to check the analytical result, we pick up the appropriate optical potential strength from domain in FIG.3(a). The right panel of FIG.7 shows the dynamics for the optical potential strength $V_s=2$ which represents the stable dynamics. In this case , one can observe that only a fraction of the condensate can oscillate in space and time. Moreover, the essential condition for dynamical stability that shows oscillation deviation of the square amplitude, is not moved away from its initial value $\phi_0=1.0$~\cite{Sabari2015}. FIG.7 (left panel) shows the dynamics of 3D space-time evolution of the square amplitude $|\phi(x,t)|^2$ of the wave for optical potential strength $V_s=0$ with the rest of parameters being the  same as in FIG.3(a). The dynamics exhibits instability as one could expect for these parameters in FIG.3(a) and FIG.5(d). The square amplitude $|\phi(x,t)|^2$ drastically moves away from $\phi_0=1.0$ to $\phi_0=4.0$.  Comparing the results displayed in two panels of FIG.7, one can realize from the growth rates  that the system is in fact unstable in left panel compared to right panel and also the wave amplitude is higher for unstable mode.
In FIG.8, we show the dynamics of 3D space-time evolution of the square amplitude of the wave, $|\phi(x,t)|^2$, where values of repulsive condensate, strength of the higher-order non-linearity ($g_2$), three-body interaction ($\chi$) and optical potential strength ($V_s$), are picked from the FIG.4(e). The left panel shows unstable dynamics for $V_s=-4$, where the initial amplitude moves away from $\phi_0=1.0$ to $\phi_0=2.0$. The right panel shows that, for stable dynamics with $V_s=0$, only fraction of condensate atoms do oscillate and, the deviation of initial amplitude is $\phi_0=1.0$ to $\phi_0=1.2$. All these results are well matched with analytical results.

\section{Conclusion}
\label{sec6}
In this paper, we have derived the modulational instability condition (see Eq.(\ref{Jac26})) for BECs trapped in an optical lattice potential. There arises corrective terms, as compared to previous available data (see ref.\cite{OlivierJOSAB2020}), due to the amount of OL potential energy.  First analytically, through a variational approach, the MI is analysed and predictions are built over the shrinking  of thickness for stability bandwidth. These results match with reality, especially in the context  of  BECs on the surface of atomic chips and in atomic waveguides involving a strong compression of the traps thereby  appreciably enhancing the density of the BECs. The nature of the condensate (attractive and repulsive) is nonsensitive to the effects of the MI criteria with OL potential.
Secondly, after trying analytical resolution of the modified GP equation through approximation method (TDVA), we obtained the ordinary differential equations in various perturbation variables. We have then analyzed in detail the role of higher-order interactions on the modulational instability of BECs immersed in OL potential through effective potential plots. For instance, we have made prominent predictions over feasible insight into the dynamics of high density BECs: When higher-order interactions are not taken into account, the MI may appear or not, depending on the strength and sign of optical lattice applied to the condensate. But in the presence of higher-order interaction, while its strength is controllable by the Feshbach resonance techniques, we have been able to suppress unstable zones thereby annihilating the modulational instability phenomena in the system. These semi analytical results  are then supported by  direct numerical integrations .  

\begin{appendix}
\appendix
\section{Derivation of variational parameters $a(t)$ and $b(t)$}

The MI-motivated trial wavefunction in Eq. (\ref{Jac8}) is substituted into the Lagrangian density in Eq. (\ref{Jac7})  and the effective Lagrangian is calculated by integrating the Lagrangian density as
\begin{align}
L_{eff}=\int \mathcal{L}  \,\,dx \nonumber
\end{align}
 But here, we consider an annular (one-dimensional) geometry, which imposes periodic boundary conditions on the wave function $\psi(x, t)$ and integration limits $0 \leqslant x < 2\pi$. This causes the quantization of the wave numbers, i.e. $k,
q = 0,\pm1,\pm2,\pm3, . . . $. In this new geometry, calculating the effective Lagrangian yields
\begin{align}
\begin{split}
&L_{\mathrm{eff}} = \pi  \Big\{-V_s \left(A_0^2+a_1^2+a_2^2\right) -g \Big[2A_0^2(a_1^2+a_2^2)-A_0^4\nonumber\\ & +4a_1^2a_2^2+4A_0^2a_1a_2 \cos(b_1+b_2)+a_1^4+a_2^4 \Big]+\eta \Big[8q^2a_1^2a_2^2 \nonumber\\ & 
+4A_0^2q^2a_1a_2 \cos(b_1+b_2) +2A_0^2q^2(a_1^2+a_2^2) \Big] -\chi\Big[\frac{2}{3}(a_1^6+a_2^6)\nonumber\\ & +4A_0^4(a_1^2+a_2^2)+6A_0^2(a_1^4+a_2^4)
+6a_1^2a_2^2(a_1^2+a_2^2) +\frac{4}{3}A_0^6 \nonumber\\ &+4A_0^2a_1a_2 \cos(b_1+b_2)(2A_0^2+3a_1^2+3a_2^2) +24A_0^2a_1^2a_2^2\Big]\nonumber\\ &
+2a_1^2\left(- 2kq-q^2-\dot{b_1}\right)+2a_2^2\left(2kq-q^2-\dot{b_2}\right)\Big\} \nonumber
\end{split}
\end{align}
The expression of this effective Lagrangian is such that the pair
$\{b_1(t), b_2(t)\}$ may be interpreted as the set of generalized
coordinates of the system, while the pair $\{A_1(t), A_2(t)\}$,
with $A_1(t)=2a_1^2(t)$ and $A_2(t)=2a_2^2(t)$, gives the
corresponding momenta. The Hamiltonian of the system is expressed
as

\begin{align}
\begin{split}
 H = &\, -L+\int_{-\infty}^{\infty}\frac{\mathrm{i}}{2}\left(\frac{\partial\psi}{\partial t}\psi^*-\frac{\partial\psi^*}{\partial t}\psi\right)\mathrm{d}x.\nonumber
\end{split}
\end{align}
Considering the integration limits imposed by the new geometry, we
have
\begin{align}
\begin{split}
&H = \pi  \Big\{V_s \left(A_0^2+\frac{A_1}{2}+\frac{A_2}{2}\right) +g \Big[2A_0^2(A_1+A_2)+A_0^4\nonumber\\ & +\frac{1}{4} (A_1^2+A_2^2)+2A_0^2\sqrt{A_1} \sqrt{A_2}\cos(b_1+b_2)+A_1A_2 \Big]\nonumber\\ & -\eta \Big[2A_0^2q^2\sqrt{A_1} \sqrt{A_2} \cos(b_1+b_2) +A_0^2q^2(A_1+A_2)  \nonumber\\ & +2q^2A_1A_2
\Big] -\chi\Big[\frac{1}{12}(A_1^3+A_2^3) +2A_0^4(A_1+A_2)\nonumber\\ & +\frac{3}{2}A_0^2(A_1^2+A_2^2)
+\frac{3}{4}A_1A_2(A_1+A_2) -\frac{4}{3}A_0^6 +6A_0^2A_1A_2 \nonumber\\ &+A_0^2\sqrt{A_1} \sqrt{A_2} \cos(b_1+b_2)(4A_0^2+3A_1+3A_2) \Big]\nonumber\\ &
+A_1\left(2kq+q^2+k^2\right)+A_2\left(-2kq+q^2+k^2\right)+2A_0^2k^2\Big\} \nonumber
\end{split}
\end{align}

In order to derive the evolution equations for the time-dependent parameters introduced in Eq.~(\ref{Jac8}) , we use the corresponding Euler-Lagrange equations based on the variational effective Lagrangian $L_{\mathrm{eff}}$. In the generalized form, these equations read
\begin{align}
\begin{split}
 \frac{\mathrm{d}}{\mathrm{d}t} \left( \frac{\partial  L_{\mathrm{eff}}}{\partial \dot \xi_i}\right)-\frac{\partial  L_{\mathrm{eff}}}{\partial \xi_i}  = 0 \nonumber,
\end{split}
\end{align}
where $\xi_i$ and $\dot{\xi_i}$ are, respectively, the generalized coordinate and corresponding generalized momentum. Hence, the evolution equation corresponding to the variational parameter $a_1$ is
\begin{align}
\begin{split}
\frac{\partial a_1}{\partial t} =
&\,\left[q^2\,\eta-g-\chi(2A_0^2-3a_1^2-3a_2^2)\right]A_0^2\,a_2\,\sin(b_1+b_2).\nonumber
\end{split}
\end{align}
For the parameter $b_1$ ,the evolution equation reads
\begin{align}
\begin{split}
\frac{\partial b_1}{\partial t} &=
-2kq-q^2-\frac{V_s}{2}-g\left(A_0^2+a_1^2+2a_2^2\right)\nonumber\\ & +q^2\,\eta \left(A_0^2+4a_2^2\right) 
-\chi\Big[2A_0^4+6A_0^2(a_1^2+2a_2^2)+a_1^4\nonumber\\ & +3a_2^2(2a_1^2+a_2^2)\Big] +\left(q^2\,\eta-g\right)A_0^2\,\frac{a_2}{a_1}\,\cos(b_1+b_2).\nonumber
\end{split}
\end{align}
For the parameter $a_2$, we get
\begin{align}
\begin{split}
\frac{\partial a_2}{\partial t}=
&\,\left[q^2\,\eta-g-\chi(2A_0^2-3a_1^2-3a_2^2)\right]A_0^2\,a_1\,\sin(b_1+b_2).\nonumber
\end{split}
\end{align}
and for the parameter $b_2$, the evolution equation is
\begin{align}
\begin{split}
\frac{\partial b_2}{\partial t} &=
-2kq-q^2-\frac{V_s}{2}-g\left(A_0^2+2a_1^2+a_2^2\right)\nonumber\\ & +q^2\,\eta \left(A_0^2+4a_1^2\right) 
-\chi\Big[2A_0^4+6A_0^2(2a_1^2+a_2^2)+a_2^4\nonumber\\ & +3a_1^2(a_1^2+2a_2^2)\Big] +\left(q^2\,\eta-g\right)A_0^2\,\frac{a_1}{a_2}\,\cos(b_1+b_2).\nonumber
\end{split}
\end{align}

For simplicity, we may use a variant of the MI-motivated wavefunction (\ref{Jac8}) for which
\begin{align}
\begin{split}
a_1 = a_2 = a, \,\,\,\,\,\,\,\,\,\,\,\,  \mathrm{and}
\,\,\,\,\,\,\,\,\,\,\,\,\,\, b_1+b_2=b.\nonumber
\end{split}
\end{align}
Then the coupled ordinary differential equations for $a(t)$ and
$b(t)$ are shown Eq.(\ref{Jac19})  and Eq.(\ref{Jac20}) 
\color{black}
\end{appendix}

\section{Acknowledgments}
SS acknowledges the Foundation for Research Support of the State of São Paulo (FAPESP) [Contracts 2020/02185-1 and 2017/05660-0]. RT acknowledges the Council of Scientific and Industrial Research (CSIR), the Government of India for Research Associateship (Grant No. 03(1456)/19/EMR-II). RR wishes to acknowledge the financial assistance received from  CSIR (Grant No 03(1456)/19/EMR-II).

\end{document}